# On the Role of Flexibility in Linker-Mediated DNA Hydrogels

Iliya D. Stoev,*[a] Tianyang Cao,*[b] Alessio Caciagli,[a] Jiaming Yu,[a] Christopher Ness,[c] Ren Liu,[a] Rini Ghosh,[a] Thomas O'Neill,[a] Dongsheng Liu[b] and Erika Eiser[‡a]

[a.] Cavendish Laboratory, Department of Physics, University of Cambridge, Cambridge, CB3 0HE, United Kingdom.
[b.] Department of Chemistry, Tsinghua University, Beijing, 100084, P. R. China.
[c.] Department of Chemical Engineering and Biotechnology, University of Cambridge, Cambridge CB3 0AS, United Kingdom.
*Authors with equal contribution.
[‡] Corresponding author: ee247@cam.ac.uk.

Three-dimensional DNA networks, composed of tri- or higher valent nanostars with sticky, single-stranded DNA overhangs, have been previously studied in the context of designing thermally responsive, viscoelastic hydrogels. In this work, we use linker-mediated gels, where the sticky ends of two trivalent nanostars are connected through the complementary sticky ends of a linear DNA duplex. We can design this connection to be either rigid or flexible by introducing flexible, non-binding bases. The additional flexiblity provided by these non-binding bases influences the effective elasticity of the percolating gel formed at low temperatures. Here we show that by choosing the right length of the linear duplex and non-binding flexible joints, we obtain a completely different phase behaviour to that observed for rigid linkers. In particular, we use dynamic light scattering as microrheological tool to monitor the self-assembly of DNA nanostars with linear linkers as a function of temperature. While we observe classical gelation when using rigid linkers, the presence of flexible joints leads to a cluster fluid with reduced viscosity. Using both the oxDNA model and a coarse-grained simulation to investigate the nanostar-linker topology, we hypothesise on the possible structure formed by the DNA clusters.

## Introduction

The vast potential of highly specific self-assembly of short DNA molecules into nano-sized building blocks, with free bases as sticky overhangs, has given rise to hierarchical self-assembly into finite-size origamis and space-filling DNA hydrogels for nearly half a century. [1-10] The reason for this interest is the ease, with which DNA of any short sequence can be produced. Consequently, DNA has also become an indispensable tool for molecular recognition. [11] Recently, the RNA-based CRISPER CAS 9 technology has been recognized as gene-editing tool to cleave short malfunctioning DNA sequences from genes, allowing the body to repair itself. [12] In other applications, DNA hydrogels have been developed as replacement for soft, tissue-friendly materials for their thermal [13] or pH-reversible responsiveness [14] and precise specificity in binding. Due to the rising interest in using this type of networks in nanotechnology for producing new functional materials, there is an increased demand for better understanding of their mechanical properties.

In most general terms, hydrogels are hydrophilic, polymer- or particle-based aqueous networks, which exhibit a certain degree of swelling and de-swelling properties that depend on their ionic strength. [15-17] One requirement for the formation of hydrogels is that the constituting polymer or colloid is water-soluble and able to form crosslinks. In contrast to the case of organogels, [18] charges play an important role in hydrogel formation, where the carbon backbone is usually hydrophobic and the monomers then need to display certain polarity. Buwalda *et al.* [19] provided a historic overview of the developments in hydrogel synthesis for biomedical applications. Recent advances in polymer chemistry allow the synthesis of hydrogels that remain stable under physiological conditions and can be used for embedding pharmaceutical agents in controlled drug release applications. [20]

Polymer hydrogels divide into subcategories based on whether the polymers are natural [21] or synthetic, [22] and depending on the type of cross-linking. In his review, Hoffman [15] listed many examples of biological and synthetic hydrogel formers. Here, we focus on the characterisation and elucidation of the structures formed by physical, thermally reversible hydrogels, composed exclusively of synthetic DNA.

DNA hydrogels are a particularly interesting type of physical hydrogel as they can be formed using either linear or branched building blocks. [23-24] These DNA-nanostars are held together through hydrogen bonds, which allow connectivity only at well-defined temperatures that correspond to the sequence and length of the underlying single-stranded DNA oligos, of which they are made. Understanding the phase behaviour of DNA hydrogels with varied valency is important for investigating the liquid-to-gel transition. [25-27] Both the critical temperature and the width of the gas-liquid coexistence region become smaller as the number of branching arms, i.e. valency of the building blocks, is reduced. In the limiting case of linear double strands, the coexistence region completely disappears as only arbitrarily long, single polymer chains form. This, of course, prevents the formation of a three-dimensional hydrogel.

One promising application of DNA hydrogels is their use in molecular imprinting. A comprehensive review on molecular imprinting is given by Alexander et al. [28] Crosslinked 3D polymer matrices can bind molecules of interest with the effect of forming a pattern that can be used for subsequent recognition. Once the desired network is formed, the molecule of interest is washed out, leaving behind free interstitial sites. If then the network is exposed again to a solution containing the target molecules, the latter will bind again. Thus, the polymer network has a 'memory' for a specific molecule.

The challenge is to read out the resulting memory changes. One possibility is measuring the system's changing viscoelastic properties. We demonstrated how data stitching from optical tweezers (OT), dynamic light scattering (DLS), video particle tracking (VPT) and bulk rheology allows extracting broadband viscoelasticity of a hydrogel formed solely by trivalent, Y-shaped DNA nanostars. [29] Xing et al. [30] reported the successful application of diffusing wave spectroscopy to study the microrheology of the same DNA hydrogels. In particular, Xing and co-workers showed that the system's elasticity drops by a factor 5-8 when flexible, non-binding thymine joints are introduced, while keeping the same cross-link density. Bomboi et al. [31] also showed that re-entrant melting of DNA hydrogels can be achieved. This unusual behaviour is due to the emergence of a second process that competes with gelation and eventually dominates.

In this work, we report the gel-formation of Y-shaped DNA nanostars that can only bind to each other via linear, double-stranded (ds)DNA linkers using DLS-based microrheology. We find that gelation can be completely suppressed when sufficiently long, flexible joints are placed between the rigid, double-stranded segment of the linker and its sticky ends. Using the oxDNA model and a further coarse-grained model we demonstrate that such a system rather forms a cluster fluid with significantly lower viscosity at lower temperatures.

**Experimental**

**DNA synthesis**

All DNA strands (sequences shown in Table 1) were synthesised in a BioAutomation Mermade12 synthesiser using a standard phosphoramidite synthesis protocol with 1 µmol scale. Then, single-stranded DNAs were cut from the solid phase by ammonia at 60°C and purified with high-performance liquid chromatography using a C18 column (5 µm, 9.4×30 mm, USA). Water/acetonitrile/TEAA (triethylammonium acetate buffer, 100 mM, pH = 7) was used as eluent with a flow rate of 3 mL/min. Finally, the purity of the products was analysed by polyacrylamide gel electrophoresis and matrix-assisted laser desorption/ionisation time-of-flight mass spectrometry. The samples were freeze-dried for subsequent use.

**Sample preparation**

The hybridisation and melting behaviour of all DNA building blocks was assessed using UV-visible spectroscopy and polyacrylamide gel electrophoresis. A NanoDrop 2000 was used to determine the concentration of the ssDNA suspensions, which were dispersed in 100 mM NaCl and 10 mM phosphate buffer (pH ≈ 7.4). A Thermal Cycler (Techne, TC-512, Bibby Scientific) was used for hybridising strands Y1-3 forming the Y-shape and for strands L1 and L2 forming the linear duplexes. After careful annealing at 94°C, the samples were cooled from 85°C down to 35°C at a rate of 0.02°C/s, holding for 20min at every 2°C. The hybridised linkers and Y-shapes were then mixed in a 3:2 ratio, unless stated otherwise. The final DNA concentrations were all in the range 0.5%w - 2.4%w (equivalent to 175 µM – 840 µM).

**DLS-based microrheology**

A Malvern Zetasizer Nano ZSP (633 nm HeNe laser) setup was used for the passive microrheology of all DNA hydrogel samples. We added to all DNA samples 230 nm polystyrene spheres coated with polyethylene glycol (PS-PEG, Cambridge Bespoke Colloids, UK), which served as probe particles. To ensure the detection of only single-scattering events, the setup was operated in a non-invasive backscatter (NIBS) mode with the detector being at an angle of 173°. In this setting, optimal scattering was achieved for a particle volume fraction of 0.03%, ensuring that the probe scattering dominated over direct scattering from the sample, accounting for over 90% of the signal.

**Table 1** DNA sequences used in the assembly of our building blocks and their corresponding melting temperatures.
[a] The SantaLucia and the measured values were obtained assuming a buffer solution containing in total 1 μM DNA and 100 mM added NaCl.
[b] Melting temperatures obtained from oxDNA simulations using a DNA concentration of roughly 5 μM and 200 mM NaCl added salt.
[c] Estimated values for 175 μM – 840 μM DNA concentration in buffer solution containing 100 mM NaCl.

| ssDNA | sequence | $T_m$ /°C SantaLucia[a] | $T_m$ /°C UV-vis measurements[a] | $T_m$ /°C oxDNA modelling[b] | $T_m$ /°C high DNA concentration[c] |
|---|---|---|---|---|---|
| Y1 | 5'-CGA TTG ACT CTC CAC GCT GTC CTA ACC ATG ACC GTC GAA G -3' | 51.11 | $T_{m,Y} \approx 55°C$ | $T_{m,Y} \approx 59$–$65°C$ | $T_{m,Y} \approx 62$–$65°C$ |
| Y2 | 5'-CGA TTG ACT CTC CTT CGA CGG TCA TGT ACT AGA TCA GAG G -3' | 48.44 | | | |
| Y3 | 5'-CGA TTG ACT CTC CCT CTG ATC TAG TAG TTA GGA CAG CGT G -3' | 41.84 | | | |
| L0(6) | 5'-GAG AGT CAA TCG (TTT TTT) TCT ATT CGC ATG ACA TTC ACC GTA AG-3' | 62.97 | $T_{m,L0(6)} \approx 65°C$ | $T_{m,L0(6)} \approx 65$–$71°C$ | $T_{m,L0(6)} \approx 72$–$75°C$ |
| L0(6) | 5'-GAG AGT CAA TCG (TTT TTT) CTT ACG GTG AAT GTC ATG CGA ATA GA-3' | 63.97 | | | |
| Y-L0(6) | 5'-CGA TTG ACT CTC  5'-GAG AGT CAA TCG | 40.54 | - | $T_{m,YL} \approx 52$–$56°C$ | $T_{m,YL} \approx 48°C$ |

The DNA hydrogel was prepared in a stepwise fashion directly in a disposable cuvette (ZEN0040, Malvern): we first dispersed our colloids in a known volume of phosphate buffer saline, ensuring they are distributed homogeneously over the entire sample volume. Then, we added a layer of our Y-shape stock solution, followed by another layer containing our linker stock solution at room temperature. This was repeated until we have added all Y-shapes and linkers. Performing this layer-by-layer addition, we ensured good mixing of our building blocks. Silicone oil (50 cSt, Sigma-Aldrich) was added on top of the hydrogel to prevent evaporation. The cuvettes were then sealed with parafilm to prevent evaporation or contamination from the outside.

The samples were subsequently measured using the Zetasizer with a thermostat operating between 60°C and room temperature. The measured intensities were analysed utilising a constrained regularisation (CONTIN) method. We normalised the raw intensity autocorrelation functions ($g_2(q, t)$) and then converted them into electric-field autocorrelation functions ($g_1(q, t)$) with Matlab routines, developed in our group. Here $q$ is the scattering wavevector and $t$ denotes the time lag. After conversion of $g_2(q, t)$ into $g_1(q, t)$, we used the relation $g_1(q, t) = \exp(-q^2 \langle \Delta r^2 \rangle /6)$ to obtain the mean-squared displacements MSD = $\langle \Delta r^2 \rangle$ as detailed by Stoev et al.[29] The final step of the analysis required Fourier transforming the MSDs, which are related to the complex shear modulus $G^*(\omega) = G'(\omega) + iG''(\omega)$ via the generalised Stokes-Einstein relation. $G'(\omega)$ and $G''(\omega)$ are the elastic and viscous moduli, which are functions of the angular frequency $\omega$.

**OxDNA modelling**

We use the coarse-grained oxDNA2 simulation package, implemented into LAMMPS,[32-33] which is based on experimentally measured thermodynamic properties of DNA duplexes, representing DNA melting/hybridisation very accurately. A graphics package including Ovito and Chimera was employed to visualise the configurations of the Y-shapes and linear linkers.

**Coarse-grained model**

We studied the structural and rheological properties of the hydrogel using our established coarse-grained simulation model.[34] The model is in the spirit of the Kremer-Grest bead-spring system,[35] treating the Y-shapes as molecules comprising seven Weeks-Chandler-Anderson spheres of radius $R$ and mass $m$ connected by harmonic bonds, while the linkers comprise three connected spheres. All parameters were chosen to match our experimental system and our oxDNA findings for the configurations and binding energies for the sticky overhangs. Harmonic three-body angle potentials keep the Y-shape arms rigid. The terminal spheres of the Y- and L-shapes hosted complementary attractive Lennard-Jones patches, shown in red and green in Figure 4. The two different linker flexibilities were set by constraining the patch position using a similar three-body angle potential found in the oxDNA simulations: in the flexible L6-scenario, the patches were allowed to move freely over their host particle surface; in the stiff L0-scenario, the patches were constrained.

## Results

**Characterisation of the DNA building blocks**

In this work, we tested how the viscoelastic properties of DNA hydrogels evolve when connecting Y-shaped DNA to each other through linear duplexes with complementary sticky overhangs. In particular, we studied the effect of non-binding flexible joints made of thymines (Ts) that were placed between the linear dsDNA linkers and their sticky overhangs. To this end, we first tested the purity and thermodynamic stability of the two building blocks separately using UV-vis spectroscopy and gel electrophoresis. The respective melting temperatures, listed in Table 1, were above ~65°C for the Y-shapes and the linear linkers, while the melting temperature $T_{m,YL}$ of the sticky overhangs was estimated to be $\geq$ 52°C for the high DNA concentrations and the added salt used, using SantaLucia's nearest-neighbour rules. [36] Details are given in the supporting information. It should be noted that the melting temperatures of the L0(6) linkers were only weakly influenced by the presence of the flexible T-joints. [37]

For visualisation and testing the system's melting behaviour of the building blocks and how they bind to each other, we also employed oxDNA simulations, presented in Figure 1. The complete hybridisation of the linear linkers and the Y-shapes as well as the binding between the two building blocks is shown. Using the fact that the persistence length of dsDNA is about 50 nm, corresponding to 150 base pairs (bps), the oxDNA model shows that the Y-L0 connection is rather rigid. Long simulation runs at around 25°C provided us with a histogram of angles $\theta_{L0}$ between the average axis of one arm or the Y-shape and that of the rigidly bound L0 linker. Introducing 6 non-binding thymines as flexible joint between the sticky overhang and the dsDNA rigid linker segment (linker L6) leads to a dramatic increase in the range of angles $\theta_{L6}$ between the linker and the same Y-shape arms, as shown in the histogram and image in Figure 1. This is due to the fact that the persistence length of ssDNA is about 1 nm, corresponding to 3 bps. Consequently, the T-joints in L6 provide a non-negligible probability that a single linker will bind to two arms of the very same Y-shape and thus reducing the effective number of linkers or valency available for the gel formation. [25]

**Microrheology**

We studied the thermally reversible gel-formation of Y-L0(6) mixtures using DLS-based microrheology. The measurements were performed on samples with a total DNA concentration ranging from 0.5%w to 2.4%w and a fixed 2:3 ratio between Y- and L-shapes, guaranteeing full hybridisation between all building blocks. We measured all samples at temperatures between 40°C and 60°C in 2°C steps, which we identified to be the melting region of the sticky overhangs ($T_{m,YL} \approx$ 52°C). Figure 2 summarises the extracted electric-field autocorrelation functions $g_1(q, t)$ and mean-squared displacements (MSD) for 0.8%w (280 µM) and 1.6%w (560 µM) DNA concentrations of rigid Y-L0 and Y-L6 solutions. For clarity, only 3 characteristic $g_1(q, t)$ curves above, at and below $T_{m,YL}$ are presented. All samples were initially heated to 65°C for 10 minutes to allow for full mixing of the Y- and L0(6)-shapes, then brought to 60°C and subsequently slowly cooled to 40°C, allowing the samples to equilibrate for 20 min at each measured temperature.

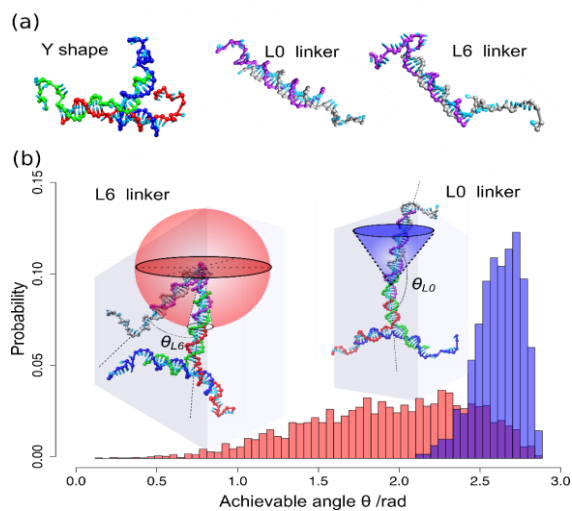

**Fig. 1** (a) oxDNA representations of the Y-shapes and linear linkers L0(6) using the sequences in Table 1. (b) oxDNA simulations of the probability that the joint between a Y-shape and a linker L0 (blue histogram) or L6 (red histogram) adopts a certain angle θ. All simulations were performed at 25°C.

## Discussion

In agreement with our previous measurements on hydrogels made only of Y-shapes with the appropriate sticky overhangs, [29,30] we observe a liquid-to-gel transition in the Y-L0 system signalling that most sticky ends are hybridised below $T_{m,YL} \approx 52°C$, as demonstrated in Figure 2a.

At around 60°C, the $g_1(q, t)$ curve for the 0.8%w concentration of the Y-L0 mixture shows a single exponential relaxation describing the unbound liquid state of the Y- and L-shapes in solution. The shift of $g_1(q, t)$ towards higher relaxation times with respect to the pure aqueous solution, containing only the probe-colloids, is simply due to the overall DNA concentration (red lines in Figure 2a). As we cool down to 44°C, the relaxation times of both concentrations (green and blue lines in Figure 2a) clearly increase faster than one would expect, assuming the Y- and L-shapes would remain unbound and only the viscosity of water would increase. Fitting the $g_1(q, t)$ curves using only the solid (0.8%w) and dashed (1.6%w) parts of the $g_1(q, t)$ curves to extract the mean-squared displacements for the different temperatures, we see that the 0.8%w Y-L0 mixture is indeed a Newtonian fluid at 60°C. Below ~ 52°C, the MSD starts showing a clear deviation from the $<\Delta r^2> = 6Dt$ to a sub-diffusive behaviour. This change coincides with the $T_{m,YL}$ found in our oxDNA simulations and marks the temperature, at which the entire sample becomes a percolating gel. [29,30] The transition from a fluid to a gel state is much more dramatic in the 1.6%w sample. It should be noted that the MSD of the 1.6%w sample shows already at 60°C a weak sub-diffusive behaviour, which is due to the increased melting temperature of the sticky overhangs between the Y- and L0-shapes. The thermodynamics of DNA hybridisation indeed shows that $T_m$ depends logarithmically on the total DNA concentration and that the width of the melting region is about 20°C. Here, we observe an increase to $T_{m,YL} \approx 54°C$, as discussed later. However, similarly to the 0.8%w sample, the system becomes a thermally reversible, viscoelastic gel.

The electric-field autocorrelation functions show a very different temperature dependence when the 6T flexible joints are introduced (Figure 2b). At 60°C, we note a similar liquid behaviour as observed in the Y-L0 mixtures, which is also confirmed through the time dependence of the MSDs. As we slowly cool the 0.8%w Y-L6 samples to 44°C we see a weak shift in $g_1(q, t)$ towards higher relaxation times, which can be accounted for by the increase in viscosity of the aqueous solvent as it is cooled. The corresponding MSDs confirm that the samples remain Newtonian fluids, although we 'switched on' the binding between the Y- and L6- shapes, while we see a clear deviation from the Newtonian behaviour in the 0.8%w Y-L0 samples. Indeed, on further cooling to room temperature we observe for both Y-L6 sample concentrations no transition to a gel phase. In fact, a sudden transition to a much more fluid sample is particularly strong in the 1.6%w Y-L6 samples and is completely different to the Y-L0 samples that always gel when cooled below $T_{m,YL}$. This trend is more visible in the corresponding MSDs, and has been observed for several concentrations (macrorheology results in the supporting information). Inspection by eye confirms the fluid nature of the Y-L6 samples and the gelled behaviour of the Y-L0 samples.

In Figure 3, we compare the elastic and loss moduli of the Y-L0 and Y-L6 systems obtained for the 1.6%w concentrations. The moduli were obtained by Fourier transforming the MSDs presented in Figure 2, and separating real and imaginary parts. Analysing $G'(\omega)$ and $G''(\omega)$ for the rigid linker hydrogel provides a plethora of information about the temperature- and frequency-dependent mechanical behaviour of our samples. First, we note the switchover of $G''(\omega)$ and $G'(\omega)$ as the temperature is lowered. A purely fluid state of Y- and L-shapes would only deliver $G''(\omega) \propto \omega$, while $G'(\omega)$ should be zero at all angular frequencies. This is true for the 0.8%w samples of both Y-L0 and Y-L6 mixtures at 60°C. However, the 1.6%w samples display a measurable elastic component at 60°C, which is nevertheless weaker than the viscous contribution in the system. At $T_{m,YL} \sim 52°C$, the $G'(\omega)$ and $G''(\omega)$ of the Y-L0 samples become similar in magnitude, overlapping with each other over a range of angular frequencies with a scaling law $G''(\omega) \propto \omega^{0.5}$. This behaviour has previously been identified as the percolation or gel point in cross-linked DNA hydrogels and polymers. [29,30,38,39] Below $T_{m,YL}$, $G'(\omega)$ starts to dominate and develop a plateau at frequencies between $10^3$ rads$^{-1}$ and $10^5$ rads$^{-1}$, which is typical for a viscoelastic gel (blue lines in Figure 3a). At lower angular frequencies, equivalent to longer times, the gel shows the characteristic scaling behaviour of a Maxwellian viscoelastic fluid with a single relaxation time $\tau_c = 2\pi/\omega_c$ denoting the cross-over between $G'(\omega)$ and $G''(\omega)$. Below $\omega_c$, the moduli scale as $G'(\omega) \propto \omega^2$ and $G''(\omega) \propto \omega$.

The plateau in $G'(\omega)$ at high frequencies can be used as an estimate of the average mesh size of the gel at that particular temperature (~44 nm), again in agreement with earlier work on Y-shapes only. [29] On further cooling, this mesh size is expected to decrease slightly as the crosslinking density increases, until full hybridisation has taken place. Cooling the sample even further would therefore not deliver higher stiffness as all possible bonds that could form have been saturated. At high frequencies, we find a scaling of $\omega^{0.75}$ for $G''(\omega)$, which is a signature of the semi-flexible nature of the DNA network. Finally, the first cross-over between $G'(\omega)$ and $G''(\omega)$ occurs at a frequency that corresponds to the longest relaxation timescale in our system. This relaxation time changes from ~ 1 ms at 52°C to ~ 2 ms at 44°C for the 1.6%w Y-L0 system.

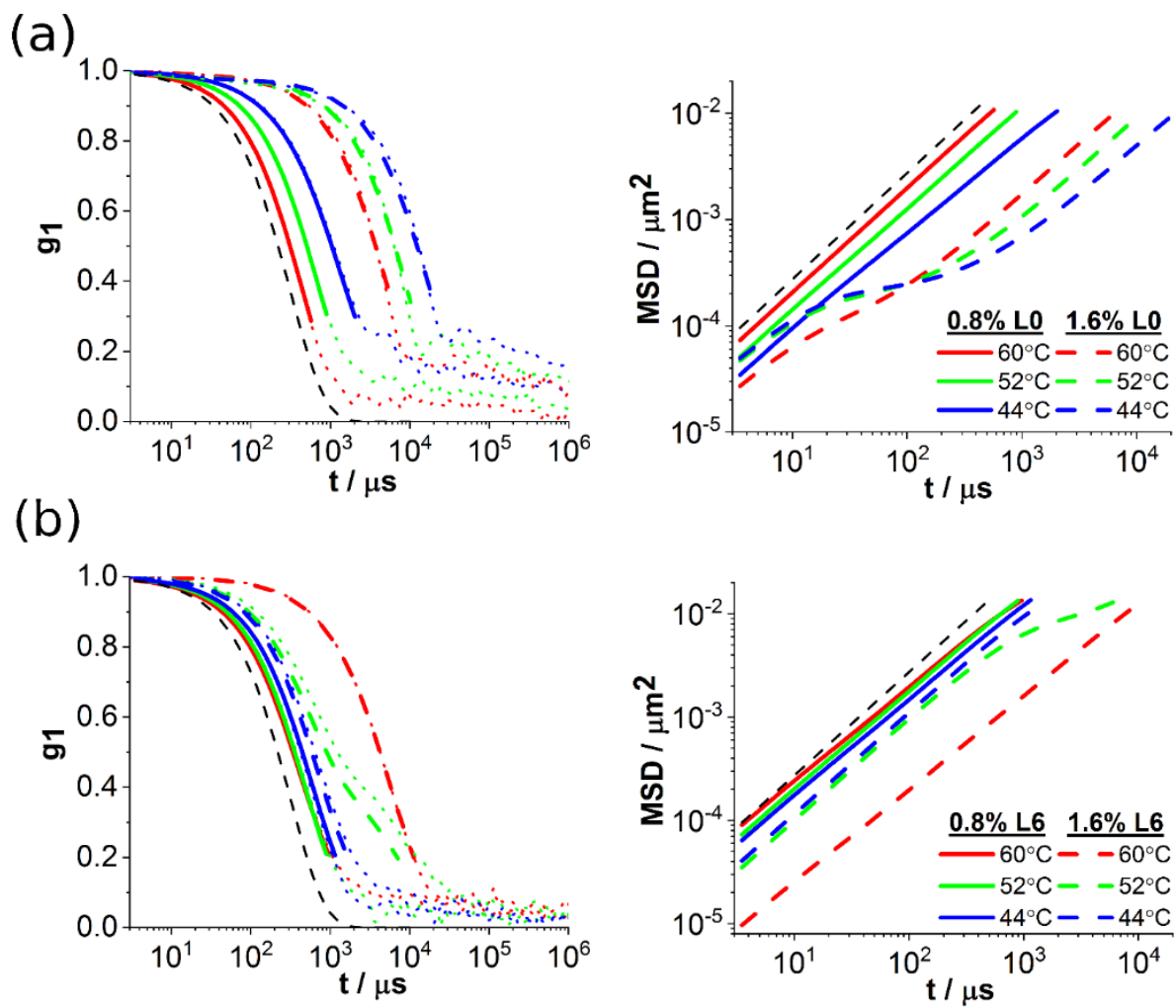

**Fig. 2** Electric-field autocorrelation functions $g_1(q, t)$ obtained from DLS measurements and the corresponding, extracted MSDs for two different concentrations of Y-shapes and linkers using (a) rigidly attached L0, and (b) L6 linkers with a flexible 6T-joint. The MSDs were calculated by fitting only the parts of the $g_1(q, t)$ curves plotted as solid (0.8%w) and dashed (1.6%w) lines. Dotted lines denote the raw $g_1(q, t)$ data as obtained from our dynamic light scattering setup. The dashed black line is the calculated relaxation curve for the same probe-particles in pure water at 60°C. All measurements were done by starting at 60°C after equilibration for 20 min at every measured temperature.

In contrast to the clear gel formation of the Y-L0 system below $T_{m,YL}$, the loss modulus of the Y-L6 system remains the dominating response over the entire range of temperatures and frequencies studied (Figure 3b). In agreement with the trends observed in the MSDs, $G''(\omega)$ decreases with temperature. Interestingly, for the 1.6%w sample we observe a low but measureable elastic contribution over the entire frequency range at 60°C. However, we see increasingly lower values and valid ranges of $G'(\omega)$ below the binding transition of the sticky arms, rendering the samples Newtonian liquids of viscosity similar to that of water.

We hypothesise that the unusual drop in the overall viscosity and the absence of any considerable elasticity below $T_{m,YL}$ in the Y-L6 system is due to a combination of two facts. Firstly, the length of the dsDNA part of the linear linkers (26 bps) is approximately the same as that of a rigid arm of the Y-shape plus the sticky ssDNA overhang when hybridised (12 + 13 bps). Secondly, the presence of the 6T flexible joint allows the linear linker to explore a wide range of bending angles θ between the rigid arms plus hybridised overhangs and the dsDNA section of the linear linker. Consequently, a single linear linker can bind to two arms of the same Y-shape with a non-negligible probability, as is illustrated and computed using our oxDNA model, shown in Figure 1. By forming these 'looped' or 'key-shaped' DNA constructs, we reduce the effective number of possible bonds or the system's valency from 3

(arms per Y-shape forming a crosslink) to a valency lower than 2. This means the system goes from a fluid phase of disconnected Y- and L6-shapes above the melting transition of the sticky overhangs to a cluster fluid below $T_{m,YL}$. While the clusters have some connectivity and elasticity, they cannot percolate as the looped ends of the clusters cannot connect to neighbouring ones. Hence, we observe an absence of a dominating bulk elastic modulus $G'(\omega)$ at low temperatures.

To test this hypothesis, we employed a coarse-grained model to probe the viscoelastic properties of the Y-L0 and the Y-L6 systems at low temperatures, where the binding part between the sticky overhangs was modelled as an attractive patchy particle with a maximal strength ($k_BT/e$ = 0.1) and an overall DNA volume fraction of about 1%. Details of this bead-spring model with stick patches are given by *Xing et al*.[34] Here, we modified our coarse-grained model to account for the flexibility between the linkers and the Y-shapes, as illustrated in Figure 4, implementing the distribution of θ for the linkers with and without the 6Ts, obtained from the oxDNA calculations.

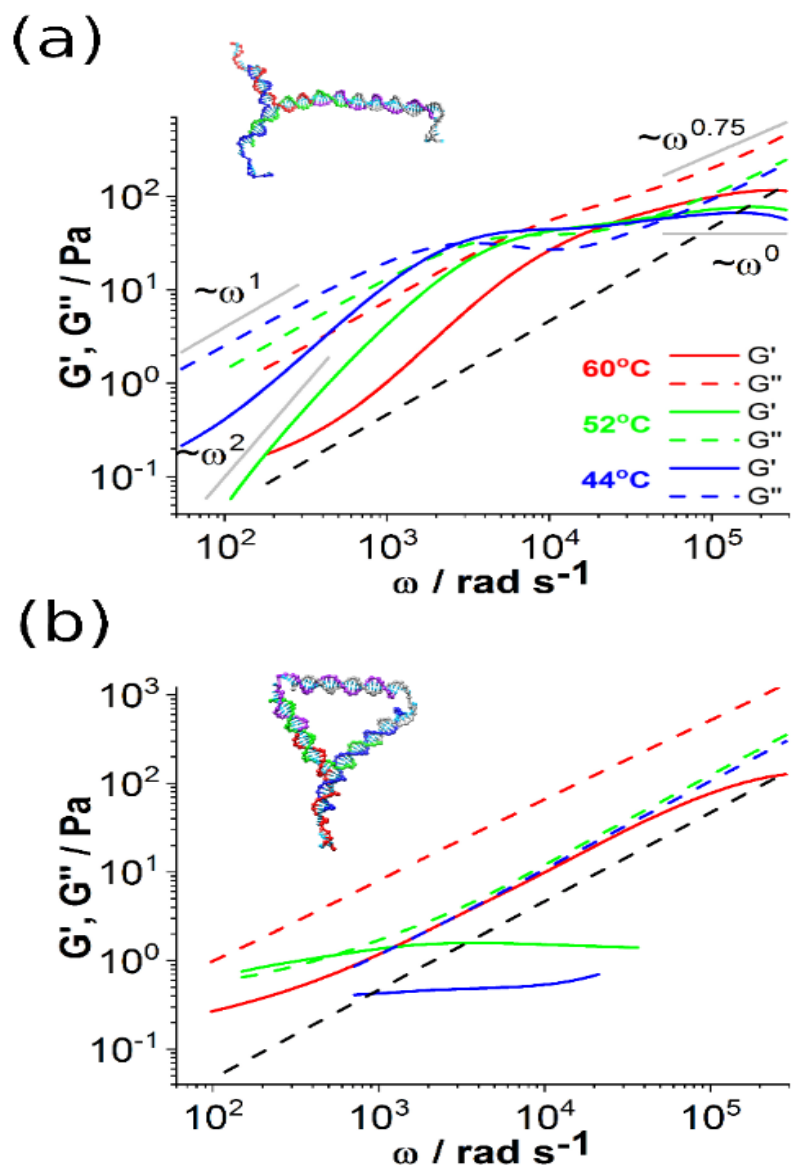

**Fig. 3** Storage and loss moduli for the 1.6%w samples made of the (a) Y-L0 system and the (b) Y-L6 system obtained for three different temperatures: above melting (60°C, red), around melting (52°C, green) and below melting (44°C, blue). The black dashed line denotes the loss modulus of water at 60°C. Grey solid lines show the scaling of G' and G'' with angular frequency. The graphic representations of how the Y-shape can bind either to L0 or L6 is obtained from oxDNA simulations.

Solving the Langevin dynamics of the coarse-grained model for high temperatures (no binding between the patches of the Y- and L-shapes; $k_BT/e = 1$) and low temperatures (fully bonded; $k_BT/e = 0.1$), we obtain $G'(\omega)$ and $G''(\omega)$ for both systems (Figure 4a). The Y-L0 system shows a clear cross-over between $G'(\omega)$ and $G''(\omega)$, with $G'(\omega)$ dominating in the gel phase. On the other hand, the Y-L6 system clearly remains liquid with $G''(\omega)$ dominating at all temperatures. Inspecting the connectivity of the two systems at low temperatures, we find that the one with rigid linkers shows a single percolating cluster, while the Y-L6 system has a clear cluster size distribution with many small clusters (Figure 4d). These results support our hypothesis that the presence of the 6T-flexible joint allows the linear linkers to bind to two arms of the same Y-shape simultaneously, thus reducing the system's effective valency to ≤ 2. Previous experimental and simulation data clearly demonstrate that a percolating equilibrium gel of DNA nanostars and patchy particles indeed only forms for valencies > 2. [25,40]

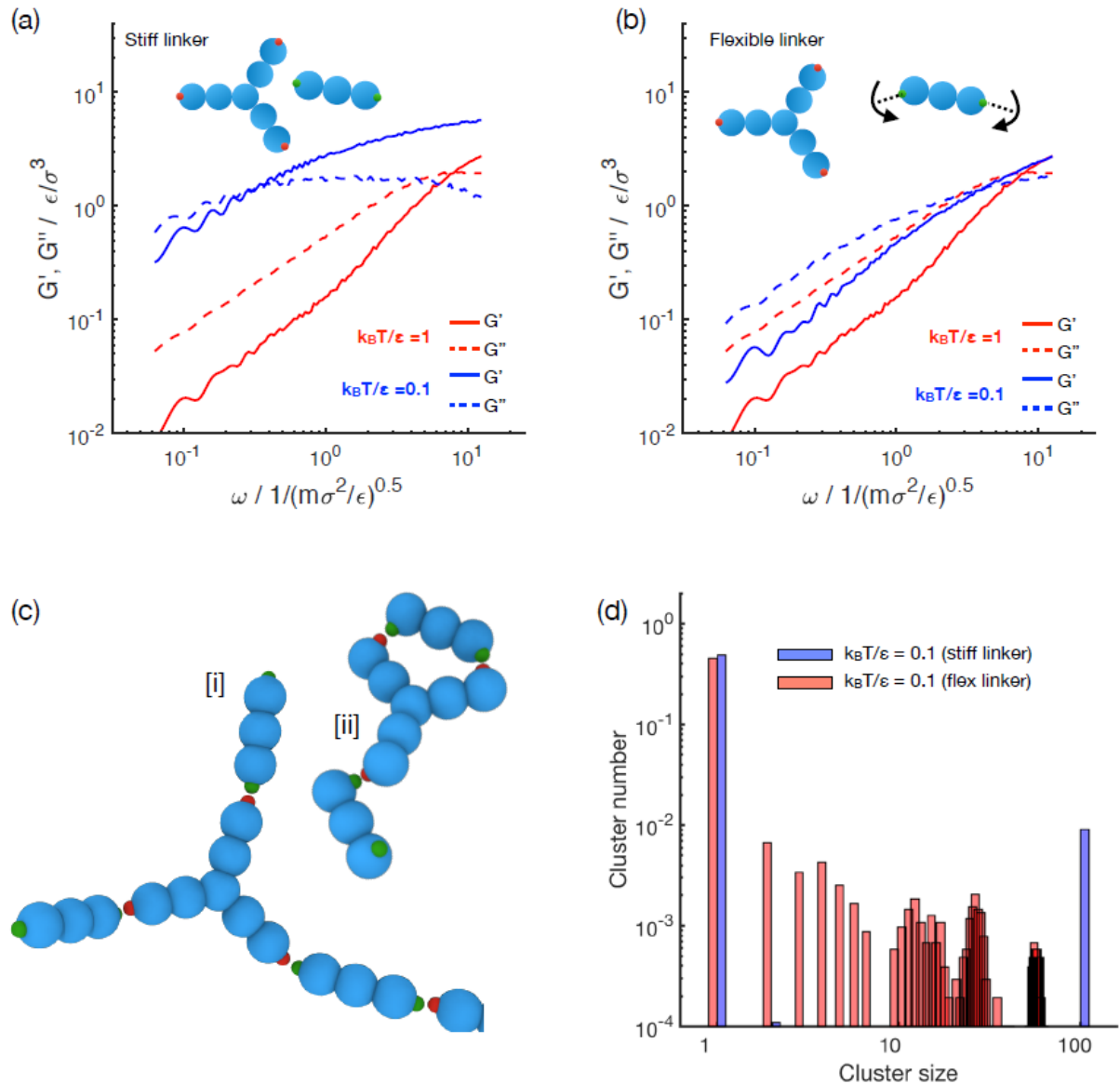

**Fig. 4**. Coarse-grained simulation model. Rheological response for mixtures of Y-shapes with (a) stiff linkers L0 and (b) flexible linkers L6. Insets show coarse-grained molecular geometry. (c) Assembled structures arising during simulation for [i] L0 linkers and [ii] L6 linkers. (d) Distribution of cluster sizes for stiff (blue bars) and flexible (red bars) linkers.

Clearly, the ability of the Y-L6 system to gel must depend also on the overall DNA concentrations and the Y:L ratio. However, our microrheology measurements on up to 2.0%w DNA concentrations show systematically a lowering of the system's viscosity at low temperatures. This observation also suggests that the preferential formation of the loops increases the translational contribution to the entropy of the emerging cluster phase.

The question that arises now is how long or short the flexible joint needs to be in order to have a considerable probability for loop formation. To this end, we measure Y-L2 and Y-L4 systems with only 2T and 4T long flexible joints, respectively (not shown here). Both showed a liquid-to-gel transition, very similar to that of the Y-L0 system. Indeed, verification with oxDNA simulations confirmed that the energy penalty to sufficiently bend these short flexible T joints outweighed the Gibbs binding free energy due to the hybridisation of the sticky overhangs. Hence, 6T signifies the minimum length for the system to fold back on itself under bending angles close to $\theta \approx 60°$.

To this end, we also tested two more Y:L6 ratios, namely 1:3 with excess linear linkers and 4:3 with excess Y-shapes; the total DNA concentration was kept constant at 1.6%w. A comparison between their MSDs and that of the optimal ratio 2:3 found previously by Xing et al. [41] is shown in Figure 5.

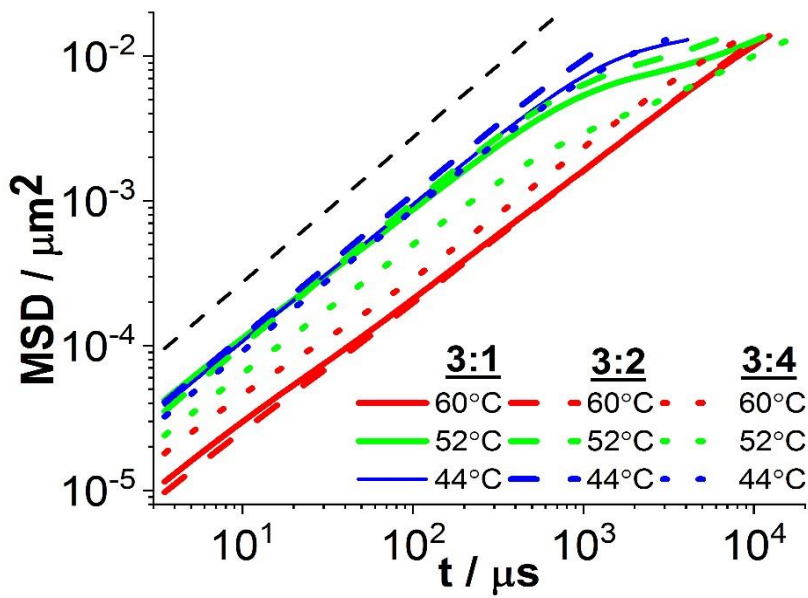

**Fig. 5** Comparing the MSDs of three different Y-L6 ratios: 1:3 (solid lines), 2:3 (dashed lines) and 4:3 (dotted lines). The measurements were performed for temperature ramps with 2°C steps between 60°C and 40°C; here, we plot only the results obtained for 60°C (red), 52°C (green) and 44°C (blue). The total DNA concentration in all samples was 1.6%w.

We chose these ratios to test the gelation ability of the Y-L6 system. In the surplus L6 system, we do expect that individual Y-shapes will rather bind to 3 L6 linkers upon cooling leading to a fully liquid solution of primarily small clusters. Increasing the Y-shape concentration beyond the optimal ratio, however, should increase the probability of the L6 linkers to bind two different Y-shapes, thus possibly leading to gelation. In our DLS measurements, all three different ratios showed roughly the same liquid-like behaviour and overall viscosity at all temperatures measured. And again, only at around $T_{m,YL}$ there is a change to sub-diffusive behaviour, visible in the MSDs, at timescales of around 1 ms. This timescale may signify simply the average lifetime of the bonds between the sticky overhangs. Interestingly, a similar timescale was observed for the cross-over of $G'(\omega)$ and $G''(\omega)$ in the gel state of the Y-L0 system. This peculiar behaviour requires further studies, which will be done in-silico, using our course-grained model (not shown here).

**Arrhenius plot for the sticky overhangs**

From the electric-field autocorrelations presented in Figure 3, we can extract the times $t_{1/2}$, at which $g_1(q, t)$ decays to half its value in the temperatures between 60°C and $T_{m,YL} \sim 52°C$. In Figure 4, these are plotted in an Arrhenius plot for different concentrations of the Y-L0 and Y-L6 systems. Assuming a relation between the half times and an energy barrier of the form $t_{1/2} = t_0 \exp(-\Delta G^{\ddagger}/(RT))$, the slopes of the curves in the Arrhenius plot give a Gibbs free activation energy of the sticky overhangs divided by the universal gas constant, $\Delta G^{\ddagger}/R$. Interestingly, the value of $\Delta G^{\ddagger}$ was surprisingly close to the values for the binding free energy $\Delta G^0$, based on Allawi and SantaLucia. [36] All predicted and measured values are given in Table 2. Notably, the progressive increase in the half times measured for increasing Y-L0 concentrations reflects the system's concentration-dependent viscosities.

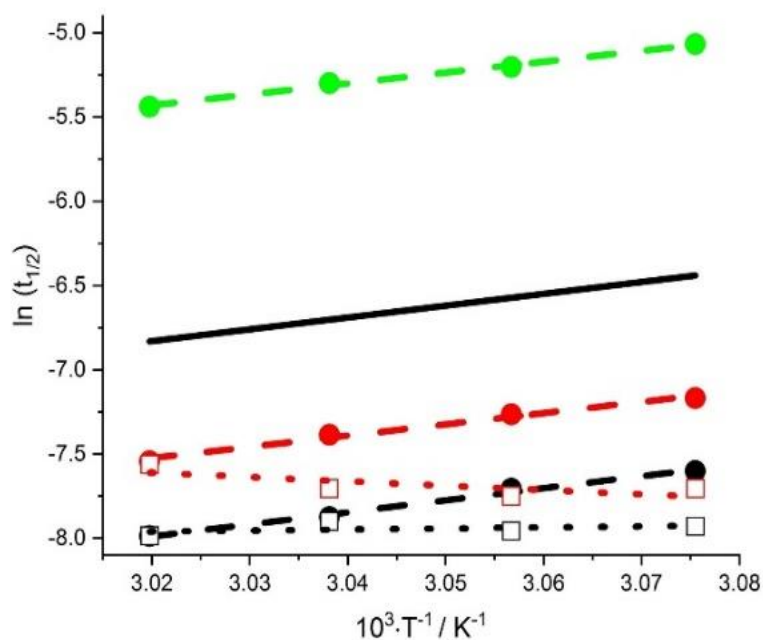

**Fig. 6** Arrhenius plot for the Y-L0 (circles) and Y-L6 samples (squares) for 0.8%w (black), 1.0%w (red) and 1.6%w (green) DNA concentrations. The slope of the solid black line is a calculated line using $\Delta G^0$ (th.) = 13.93 kcal mol$^{-1}$ for the sticky ends of the Y- and L-shapes, based on values reported by Allawi and SantaLucia, [36] and an arbitrary intercept, serving as guide to the eye.

On the other hand, the Arrhenius plots for the Y-L6 system in Figure 6 exhibit a negative gradient, implying a positive value for the activation barrier. This is equivalent to saying that the diffusing probe-particles only experience viscous drag and no bond-breaking between the sticky overhangs of the Y- and L6-shapes, which is in accordance with our experimental and simulated microrheology results.

**Effect of DNA concentration on the overall viscosity**

From the MSDs in Figure 2, we extract an apparent viscosity of the sample from the long-time behaviour at each measured temperature and plot these viscosities for both rigid and flexible linkers in Figure 7. The rigid linker hydrogel displays an increase in viscosity upon cooling, consistent with the picture of forming a transient network. In the flexible case, we observe a maximum viscosity with an associated temperature that appears to shift to lower values with increasing total DNA concentration. This behaviour implies that gelation may be favoured for higher DNA concentrations than the ones used here, as discussed above. However, within the concentration range we investigated, the gelation was impeded by another competing process. Indeed, as the samples were cooled down to 40°C, their viscosity became noticeably lower than the viscosity of the starting solution of Y-shapes and linkers. Again, as discussed above, we attribute this non-Arrhenius behaviour to the formation of a cluster phase, such that our probe-particles experienced only an aqueous solution containing a low volume fraction of soft spheres (DNA clusters).

**Table 2** Gibbs free energy extracted from the gradient of each line in Figure 6. The amount of energy necessary for breaking the hydrogen bonds between the bases in the sticky end is roughly independent of total DNA concentration, as correctly predicted by the theoretical model of Allawi and SantaLucia. [36]

| Sample | $\Delta G^{\ddagger}$ (exp.) / kcal mol$^{-1}$ | $\Delta G^0$ (theo.) / kcal mol$^{-1}$ |
|---|---|---|
| 0.8%w L0 | -13.9 ± 0.7 | -13.93 |
| 1.0%w L0 | -13.3 ± 0.7 | -13.93 |
| 1.6%w L0 | -14.0 ± 0.7 | -13.93 |
| 0.8%w L6 | +3.8 ± 0.2 | -13.93 |
| 1.0%w L6 | +4.5 ± 0.2 | -13.93 |

Figure 8 presents the variation of viscosity obtained from the long-time response in the MSDs as a function of the total DNA concentration used. In all data series, we see an increase of the viscosity with concentration. However, the temperature trend for the flexible linker fluids is reversed, as already shown in Figure 7. We fitted the data with a power-law of functional form $\eta = \eta_0 + Ac^B$, where the intercept represents the viscosity $\eta_0$ of water at the corresponding temperature [42] and $c$ stands for total DNA concentration. In the rigid linker case, the fits become less accurate with lowering the temperature due to the approach of the gelpoint, where we no longer expect a simple power-law fit to hold. The fitting parameters, prefactor $A$ and exponent $B$, were optimised using a Levenberg-Marquardt iterative algorithm [43] and are plotted in Figure 9. To our knowledge, the physical meaning of $A$ and $B$ has so far not been provided in the context of DNA hydrogels. Here, we observe a cross-over in the magnitudes of $A$ and $B$ for the rigid linker hydrogel on approaching the gelpoint, but in the flexible linker case $A$ remains lower than $B$ at all temperatures. This observation, combined with the physical appearance of the sample cuvettes and the DLS microrheology data shown in previous figures, suggests a possible relation of $A$ to $G'$ and $B$ to $G''$. A more detailed viscosity study is needed for confirming these hypothesised relations.

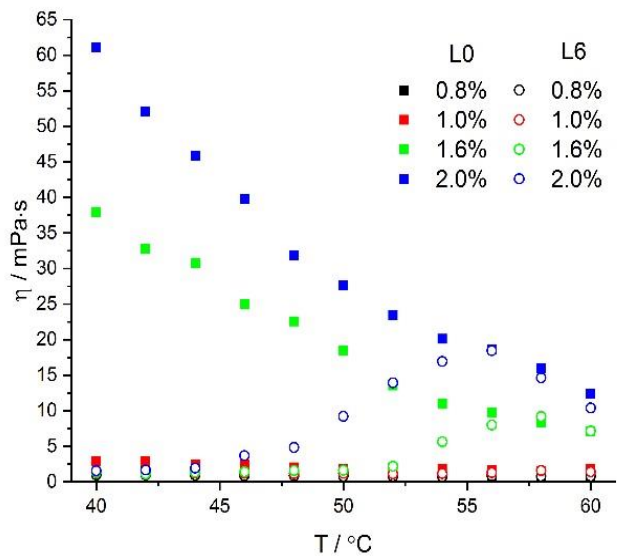

**Fig. 7** Viscosity plotted as a function of temperature for both rigid (filled squares) and flexible (empty circles) linker hydrogels. The different concentrations are labelled using the same colour coding as in Figure 6, with the addition of 2.0%w in blue. The flexible linker hydrogels appear to display a maximum in the viscosity data, which shifts to lower temperature values as the total DNA concentration increased.

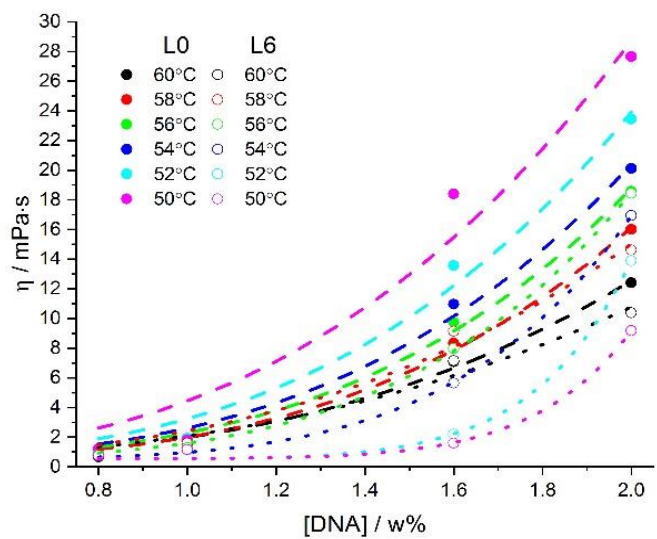

**Fig. 8** Plot of the viscosities as functions of the total DNA concentration for temperatures above and around the theoretical $T_m$. Rigid linker hydrogel data points are represented by filled circles, while data for the flexible linker fluids are displayed by empty circles. The power-law growth of each temperature series is shown as either a dashed line (L0) or a dotted line (L6).

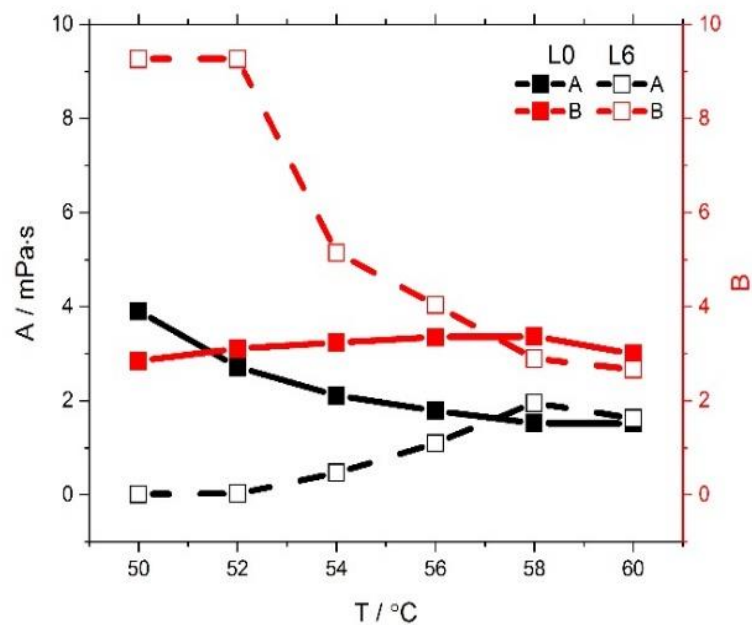

**Fig. 9** Fitting parameters *A* (prefactor) and *B* (exponent) for rigid linker hydrogels (filled squares) and flexible linker hydrogels (empty squares). Optimised values for *A* and *B* have been obtained through power-law fits utilising the Levenberg-Marquardt iterative algorithm.

## Conclusions

Here, we show that the formation of DNA hydrogels from well-defined, nanometric DNA building blocks can be tuned by introducing programmed flexibility. Using as building blocks Y-shaped DNA nanostars, all with the same sticky ssDNA overhangs, and linear duplexes with the complementary sticky ssDNA overhangs, we showed that their 2:3 mixtures can form either a thermally reversible, percolating network with substantial elastic contribution, or a cluster fluid. Which of the two phases is formed depends crucially on the introduction of sufficiently long sequences of non-binding thymines between the sticky ends and the rigid linear duplex connecting two Y-shapes. Indeed, our experimental, structural oxDNA and coarse-grained simulations prove that if the stiff arms for the Y-shapes and the linear duplex are similarly sized, a minimum of a 6T long flexible joint can lead to the linker binding to two arms of a Y-shape simultaneously, thus reducing the overall system valency for binding. The resulting cluster fluids show considerably lower viscosities at room temperature than the ones, at which all building blocks are disconnected from each other. In particular, our experiments and oxDNA calculations show that reducing the flexible joint to 4Ts or 2Ts would require a substantial increase in the bending elasticity penalty of the DNA arms to form such a cluster phase – thus, gelation is preferred. To conclude, the precise design of the DNA building blocks is crucial when developing DNA-based functional hydrogels.

## Conflicts of interest

There are no conflicts to declare.

## Acknowledgements

We would like to acknowledge Dr Zhongyang Xing for her help in the interpretation of our microrheology data. I. D. S. thanks EPSRC (grant no. 1805384) for financial support. T. C. and D.L. acknowledge support by National Natural Science Foundation of China (NNSFC, 21534007 and 21821001). A. C. and E. E. thank the Winton Program for Sustainable Physics. J. Y. and R. L. acknowledge help from Cambridge Trust and China Scholarship Council (CSC). C. N. acknowledges the Maudslay-Butler Research Fellowship at Pembroke College, Cambridge, for financial support.